\documentclass{aa}
\usepackage{txfonts,graphicx}
\usepackage{natbib}
\bibpunct{(}{)}{;}{a}{}{,} 
\usepackage{epsfig}
\newcommand{\kms}{${\rm km \, s^{-1}}$}
\newcommand{\teff}{$T_{\rm eff}$}

\begin{document}
\title {{\sc\Large Letter to the Editor}\\[5mm]
LMC origin of the hyper-velocity star HE\,0437$-$5439\thanks{Based on
observations collected at the European Southern Obser\-vatory, Paranal, Chile
(DDT proposal 276.B-5024).}}
\subtitle{
Beyond the supermassive black hole paradigm}
\author {
N. Przybilla\inst{1} \and
M. F. Nieva\inst{1} \and
U. Heber\inst{1} \and
M. Firnstein\inst{1} \and
K. Butler\inst{2} \and
R. Napiwotzki\inst{3} \and
H. Edelmann\inst{1}}
\offprints{przybilla@sternwarte.uni-erlangen.de}
\institute{Dr. Remeis-Sternwarte Bamberg, Sternwartstr. 7, D-96049 Bamberg,
Germany \and Universit\"atssternwarte M\"unchen, Scheinerstr. 1, D-81679
M\"unchen, Germany \and Centre for Astrophysics Research, University of Hertfordshire,
College Lane, Hatfield AL10 9AB, UK
}

\date{Received... ; accepted ... }
\abstract
{Hyper-velocity stars move so fast that only a supermassive black hole
(SMBH) seems to be capable to accelerate them. Hence the Galactic centre (GC) is their
only suggested place of origin. Edelmann et al.~(2005) found the
early B-type star HE\,0437$-$5439 to be too
short-lived to have reached its current position in the Galactic halo if
ejected from the GC, except if being a blue straggler star. 
Its proximity to the Large Magellanic Cloud (LMC) suggested an origin from this galaxy.}
{The chemical signatures of stars at the GC are significantly
different from those in the LMC. Hence, an accurate measurement of the abundance
pattern of HE\,0437$-$5439 will yield a new tight constraint on the place of
birth of this hyper-velocity star.}
{High-resolution spectra obtained with UVES on the VLT are analysed using state-of-the-art
non-LTE modelling techniques.}
{We measured abundances of individual elements to very high accuracy in
HE\,0437$-$5439 as well as in two reference stars, from the LMC and the solar
neighbourhood, respectively. The abundance pattern is not consistent
at all with that observed in stars near the GC, ruling our an origin from the GC.
However, there is a high degree of consistency with the LMC abundance pattern.
Our abundance results cannot rule out an origin in the outskirts of the
Galactic disk. However, we find the life time of HE\,0437$-$5439 to be more
than~three times shorter than the time of flight to the edge of the disk,
rendering a Galactic origin unlikely.}
{Only one SMBH is known to be present in Galaxy and none in the LMC.
Hence the exclusion of an GC origin challenges the SMBH paradigm. We conclude
that there must be other mechanism(s) to accelerate stars to hyper-velocity
speed than the SMBH. We draw attention to dynamical ejection from dense massive
clusters, that has recently been proposed by Gvaramadze et al.~(2008).}
\titlerunning {LMC origin of the hyper-velocity star HE\,0437$-$5439}
\authorrunning {Przybilla et al.}
\maketitle
\section{Introduction}
Stars moving faster than the Galactic escape velocity were~first
predicted to exist by \citet{hills88}. The first such hyper-velocity
stars~(HVSs) were discovered serendipitously only recently
\citep[][E05: \object{HE\,0437$-$5439} at 723\,km\,s$^{-1}$ 
heliocentric radial velocity]{brown05,hirsch05,edelmann05}.
A systematic search for such objects has resulted in the discovery of seven
additional HVS up to now \citep[see][]{brown07}.

\cite{hills88} predicted that the tidal disruption of a binary by a
supermassive black hole (SMBH) could lead to the ejection of
stars with velocities exceeding the escape velocity of our Galaxy. The Galactic
centre (GC) is the suspected place of origin of the HVSs as it hosts a SMBH.

Both, the moderate rotational velocity ($v\sin i$\,$=$\,54\,\kms)
and the chemical composition (consistent with solar abundances within a
factor of a few from a LTE analysis of a spectrum with S/N\,$\approx$\,5)
were considered evidence for a main sequence nature of HE\,0437$-$5439 (E05). 
This put the star at a distance of $\approx$60\,kpc.

Numerical kinematical experiments were carried out to trace the trajectory
of HE\,0437$-$5439 from the GC to its present location in the Galactic halo.
However, its travel time (100\,Myrs)
was found to be much longer than its main sequence lifetime
($\approx$\,25--35\,Myrs), rendering a GC origin unlikely, though possible if
HE\,0437$-$5439 would be a blue straggler star.
Alternatively the star could have originated from the Large Magellanic Cloud
(LMC) as it is much closer to this galaxy (18\,kpc) than to the~GC.

An accurate determination of its elemental abundance pattern should allow to
distinguish between an origin in the LMC or in the GC because their chemical
compositions differ distinctively. HE\,0437$-$5439 is by far the brightest HVS known
\citep[$V$\,$=$\,16.36,][]{bonanos08} and therefore suited best
for a detailed analysis.
Hence we obtained high-resolution, high S/N spectra of HE\,0437$-$5439 at the ESO VLT
with UVES and performed a quantitative spectral analysis using
state-of-the-art non-LTE modelling techniques.

\section{Observations and quantitative non-LTE analysis}
\begin{figure*}
\centering
\resizebox{.78\hsize}{!}{\includegraphics{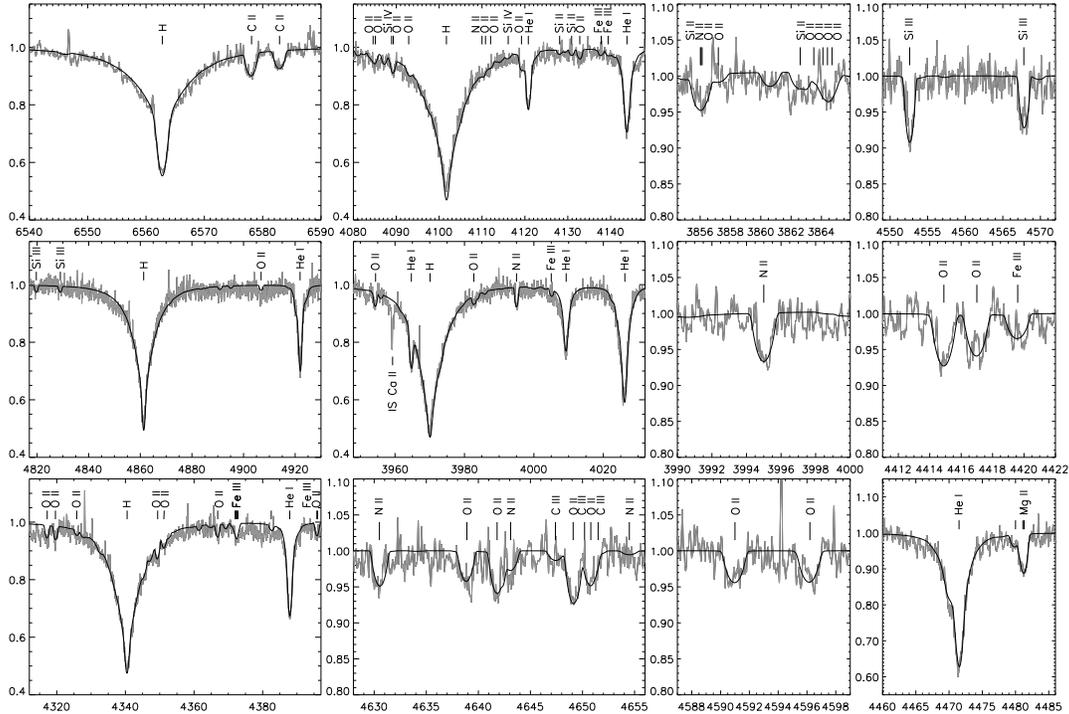}}
\caption{Comparison of spectrum synthesis for HE\,0437$-$5439 (full line) 
with observation (grey). The main spectral features are identified.
Weak interstellar \ion{Ca}{ii} H \& K lines are due to the Galactic
foreground, indicative for only a small amount of reddening.}
\label{fits}
\end{figure*}

Nine spectra of HE\,0437$-$5439 with a total integration time of $\sim$4\,h
were obtained on January 12, 2006, covering the range from 3750 to 4950\,{\AA} and
5700 to 9400\,{\AA} at a resolving~power of $R$\,$\approx$\,35\,000.
The data reduction followed procedures described by \citet{koester01}.
We took a spectrum of the DC white dwarf \object{WD0123$-$262}, such that a
reliable continuum normalisation in the blue could be achieved, a region
containing many broad Balmer lines.
Moreover, the method removes fringes in the red reliably, improving
considerably on the standard UVES data reduction pipeline. A peak S/N of $\approx$80
per resolution element is measured for the coadded spectrum in the blue.

Two comparison stars with similar atmospheric parameters are 
considered: the LMC object \object{NGC\,2004-D15}
\citep[K02]{korn02} and \object{HR\,3468} in the solar neighbourhood
\citep[NP08]{NiPr08}. The same data reduction procedures as used for
HE\,0437$-$5439 are applied to the UVES spectrum of NGC\,2004-D15, obtained
from the ESO archive. A comparison with the LMC star allows for a
consistency test with the LMC baseline abundances determined from main
sequence B-type stars in previous studies \citep[e.g. K02;][]{Trundleetal07}.

The quantitative analysis of the 3 stars was carried out following the methodology
discussed by \citet{NiPr07} and NP08. In brief, non-LTE line-formation
calculations were performed on the basis of line-blanketed model atmospheres,
using updated versions of {\sc Detail} and {\sc Surface} \citep{gid81,
butgid85}. State-of-the-art model atoms were employed~-- H: \citet{PrBu04};
\ion{He}{i}: \citet{Pr05}; \ion{C}{ii/iii}: \citet{NiPr06}, NP08;
\ion{N}{ii}: \citet{PrBu01}; \ion{O}{i}: \citet{Pretal00}; \ion{Mg}{ii}: \citet{Pretal01};
\ion{Fe}{iii} as adopted by \citet{Moetal07}; an updated model for
\ion{O}{ii}, originally by \citet{BeBu88}. Our extensions and improvements of the original
\ion{Si}{ii/iii/iv} model atom from \citet{BeBu90} were crucial for our
analysis as this ionization equilibrium is our primary temperature indicator.
Tests with high-S/N spectra of Galactic stars showed that the models allow the observed
spectra for these elements to be reproduced with high accuracy and therefore
atmospheric parameters and elemental abundances to be determined with small
uncertainties.

\begin{table}
\caption{Stellar parameters of the programme stars \& elemental abundances
$\varepsilon(X)=\log\,(X/{\rm H})+12$ (number of lines
analysed in parentheses)}
\vspace{-2mm}
\label{tab_results}
\setlength{\tabcolsep}{2mm}
\begin{tabular}{lrrr}
\hline
\hline
      & HE\,0437$-$5439 & NGC2004-D15 & HR\,3468\\
      \hline
\teff\,(K)         & 23\,000$\pm$1000 & 21\,500$\pm$1000 & 22\,900$\pm$400\\
$\log g$\,(cgs)    &    3.95$\pm$0.10 &    3.70$\pm$0.10 &    3.60$\pm$0.05\\
$\xi$\,(km/s)      &       1$\pm$1    &       1$\pm$1    &       5$\pm$1\\
$v \sin i$\,(km/s) &      55$\pm$2    &      48$\pm$3    &      11$\pm$2\\
$M$/M$_{\sun}$     &     9.1$\pm$0.8  &     9.5$\pm$0.8  &    11.6$\pm$0.5\\
Age\,(Myr)         &      18$\pm$3    &      23$\pm$1    &    14.5$\pm$0.5\\  
\hline
$\varepsilon$(\ion{C}{ii})   & 8.13$\pm$0.12\,~~(3)  & 8.05$\pm$0.10\,~~(1)  & 8.36$\pm$0.10\,(17)\\
$\varepsilon$(\ion{C}{iii})  &             {\ldots}  &             {\ldots}  & 8.47$\pm$0.04\,~~(2)\\
$\varepsilon$(\ion{N}{ii})   & 7.58$\pm$0.04\,~~(7)  & 7.15$\pm$0.06\,~~(4)  & 8.10$\pm$0.08\,(54)\\
$\varepsilon$(\ion{O}{i})    & 8.70$\pm$0.10\,~~(1)  &             {\ldots}  & 8.82$\pm$0.02\,~~(4)\\
$\varepsilon$(\ion{O}{ii})   & 8.71$\pm$0.07\,(40)   & 8.52$\pm$0.09\,(31)   & 8.80$\pm$0.09\,(45)\\
$\varepsilon$(\ion{Mg}{ii})  & 7.40$\pm$0.10\,~~(1)  & 7.25$\pm$0.10\,~~(1)  & 7.67$\pm$0.03\,~~(2)\\
$\varepsilon$(\ion{Si}{ii})  & 7.20$\pm$0.10\,~~(3)  & 7.37$\pm$0.15\,~~(3)  & 7.47$\pm$0.09\,~~(7)\\
$\varepsilon$(\ion{Si}{iii}) & 7.26$\pm$0.05\,~~(4)  & 7.30$\pm$0.12\,~~(3)  & 7.43$\pm$0.06\,~~(5)\\
$\varepsilon$(\ion{Si}{iv})  &             {\ldots}  &             {\ldots}  & 7.52$\pm$0.05\,~~(2)\\
$\varepsilon$(\ion{Fe}{iii}) & 7.47$\pm$0.10\,~~(3)  & 7.20$\pm$0.13\,~~(3)  & 7.40$\pm$0.11\,(37)\\
\hline
\end{tabular}
\end{table}

The atmospheric parameters (\teff, $\log g$) were determined from metal
ionization equilibria (mainly \ion{Si}{ii/iii/iv}, but also
\ion{O}{i/ii} and \ion{C}{ii/iii}, where available) and the
Stark-broadened Balmer lines. Microturbulence $\xi$ was
derived in the standard way by demanding that the line
equivalent widths within an ion be independent of abundance.
Elemental abundances were determined
from $\chi^2$-minimi\-sation of fits to individual metal line~profiles.

\begin{figure}
\centering
\resizebox{.9\hsize}{!}{\includegraphics{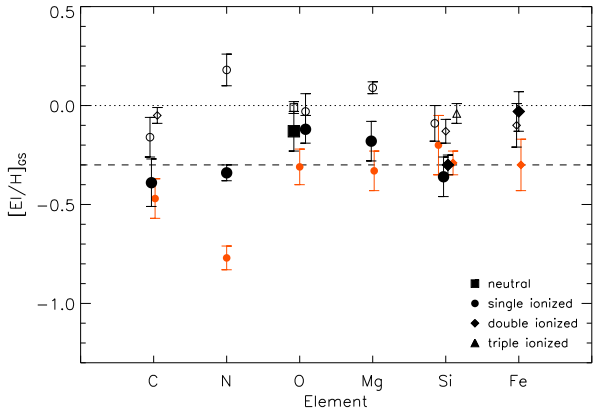}}\\[-3.5mm]
\caption{Metal abundances in the three programme stars
(Table~\ref{tab_results}), rela\-tive to the
solar standard \citep{GrSa98}: HE\,0437$-$5439 (black symbols),
NGC\,2004-D15
(red symbols) and HR\,3468 (open symbols). Ionic species are encoded according
to the legend. Solar and half-solar values are indicated by dotted and
dashed lines, respectively.}
\label{abus}
\end{figure}

Basic stellar parameters and metal abundances
for the programme stars are listed in Table~\ref{tab_results}.
Masses and evolutionary lifetimes were determined by comparison with
evolutionary tracks from \citet{schaller92} and \citet{schaerer93}, for solar 
and LMC metallicity, respectively. Note that the actual metallicity of
HE\,0437$-$5439 is $Z$\,$\approx$\,0.013, which gives a significantly lower
age (18\,Myr) than found by E05 (25--35\,Myr). The corresponding distance to
HE\,0437$-$5439 is 61$\pm$9\,kpc, in perfect agreement with E05.

The uncertainties in the atmospheric parameters were constrained by the quality of
the simultaneous fits to all diagnostic indicators: all hydrogen and helium
lines, and multiple metal ionization equilibria. Such a good match has been reported never
before in B-star analyses. The helium abundance in the three stars was found to be 
consistent with the solar value.
Standard deviations $\sigma$ for metal abundances were calculated from
the individual line abundances in an ion. The derived uncertainties are extremely low
for B-stars analyses. We adopted $\sigma$\,$=$\,0.10\,dex for ions with only one observed line.
In addition, systematic errors need to be considered because of uncertainties in the stellar
parameters, atomic data and the quality of the spectra. In fact,
the precision of the analysis is limited mostly by the noise level of the spectra.
A laborious procedure to minimise systematics has been developed by NP08,
which allows us to estimate the systematic
uncertainties in elemental abundances to be 0.10\,dex for HR\,3468 and 0.15\,dex for the
other two stars. This significant improvement with respect to previous B
star analyses can be obtained only when major
sources of systematic errors are eliminated (NP08).
The resulting synthetic spectrum is compared with
observation of HE\,0437$-$5439 in Fig.~\ref{fits}, for many strategic spectral regions.
The match between theory and observation is excellent within the S/N
limitations.
The fits obtained for the other two~stars are of similar quality.

Abundance patterns for the three programme stars are
visualised in Fig.~\ref{abus}, relative to the solar standard
\citep{GrSa98}\footnote{Similar conclusions are drawn in the following
if the revised solar abundances of \citet{GAS07} are used.}. 
Note the small line-to-line-scatter and the agreement of abundances for
different ions of the elements,
{\em simultaneously} for all ionization equilibria. This can be achieved
only when the stellar parameters are constrained well and highly reliable
model atoms are employed in the non-LTE calculations.
Metal ionization equilibria show a much
higher sensitivity to parameter variations than the H/He 
lines alone. This explains part of the parameter offset with regard to the
LTE analysis of HE\,0437$-$5439 by E05. We derived a somewhat
lower \teff~and $\log g$ for NGC\,2004-D15 than K02, though
within the mutual~uncertainties.

Elemental abundances in NGC\,2004-D15 are about half solar, except for
nitrogen, which is in general agreement with the baseline metallicity derived in
previous studies of early-type stars in the LMC.
Nitrogen is known to have a low pristine abundance in the LMC (e.g. K02).
We thus confirm NGC\,2004-D15 to be essentially
unaffected by mixing with CN-cycled material.
Abundances in the Galactic star HR\,3468 are
near-solar, except for enriched N. The abundances in
HE\,0437$-$5439 tend to be intermediate, with a solar N/C ratio.

\section{Constraints on the origin of HE\,0437$-$5439 from its chemical signature}
We shall now compare the abundance pattern of HE\,0437$-$5439 to the
chemical signatures of the two suggested places of origin, the Galactic centre
on the one hand and the LMC on the~other.

\begin{figure}
\centering
\resizebox{.9\hsize}{!}{\includegraphics{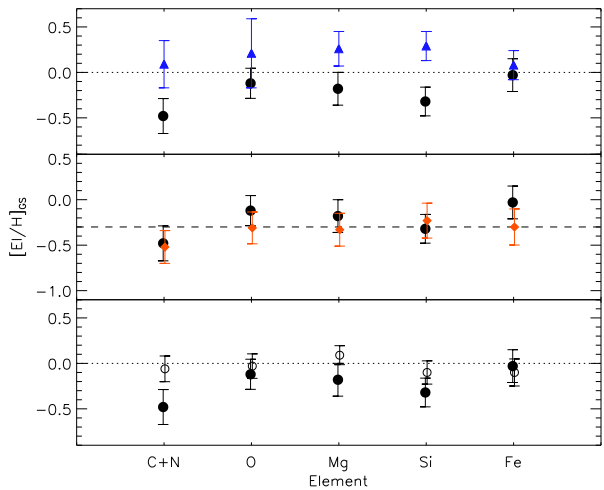}}\\[-2mm]
\caption{Comparison of metal abundances in HE\,0437$-$5439 (black dots) with
a sample of stars close to the GC (blue triangles), upper panel,
the LMC reference star NGC\,2004-D15 (red diamonds), middle pa\-nel, and
the solar neighbourhood reference star HR\,3468 (circles), low\-er panel.
Error bars account for statistic and systematic uncertainties.}
\label{abuscomp}
\end{figure}

\paragraph{The Galactic centre.} A comparison of the abundance pattern of
HE\,0437$-$5439 to that of a sample of stars near the GC
\citep{cunha07,najarro08} is depicted in Fig.~\ref{abuscomp} (upper panel), for average
values over all ions of an element. The sum of carbon and nitrogen is
considered to remove signatures of mixing with nuclear-processed matter
from the individual abundances (as catalysts, their total number is
conserved in the CN-cycle). The $\alpha$-elements oxygen, magnesium, silicon, 
and C$+$N are super-solar and enhanced relative to  iron in the GC sample. In
HE\,0437$-$5439, however, they are subsolar and depleted with respect to iron.
This rules out HE\,0437$-$5439 to be of GC origin.\\[-8mm]
\paragraph{LMC.} A comparison of the abundance pattern in HE\,0437$-$5439
with the LMC reference star is made in the middle panel of Fig.~\ref{abuscomp}.
Both patterns are very similar as all
error bars overlap. There is a tendency for the abundances of
HE\,0437$-$5439 to be slightly higher than in NGC\,2004-D15 
(from 0.04\,dex for C$+$N to 0.27\,dex for~Fe, with an exception in Si).
These abundances are fully
consistent with today's knowledge of the abundance
scatter within the LMC \citep[e.g.][]{LuLa92,hill95},
consistent with an origin of HE\,0437$-$5439 in this galaxy.\\[-8mm]
\paragraph{Galactic disk.} We are reluctant to accept the LMC origin before
having discussed other options for a Galactic disk origin. From
Fig.~\ref{abuscomp} (lower panel) it can be seen that HE\,0437$-$5439 and the
solar neighbourhood star HR\,3468 have similar abundances to
within error limits, when compared on an element-to-element basis. However,
there is a tendency for all elements except iron to be less abundant in
HE\,0437$-$5439, by 0.09\,dex (O) to 0.27\,dex (Mg). Such an
abundance pattern would restrict the place of birth of HE\,0437$-$5439
to outside the solar circle because of the Galactic abundance gradient.
Consequently, we cannot rule out an origin in the outskirts of the Galactic disk
from the chemical signature alone.

\section{Kinematics revisited} 
\citet{edelmann05} suggested the LMC as place of origin solely because of 
the discrepancy between evolutionary life time and 
time of flight. We have recalculated the kinematics on the basis of the 
significantly reduced age of HE\,0437$-$5439, following the procedure of E05. 
Such an estimate has also to account for the large
space motion of the LMC \citep[e.g.][]{piatek08} of about
500\,pc/Myr in direction east-north-east. The total distance to travel by
HE\,0437$-$543 after ejection $\sim$18\,Myr ago is therefore
$\sim$19$^{+7}_{-3}$\,kpc. Hence, the total ejection
velocity should amount to $\sim$1000\,km\,s$^{-1}$,
if originating in the kinematic centre of the LMC.
Accounting for the extension of the LMC, ejection velocities may
vary by $\sim$100\,km\,s$^{-1}$.

Flight times from the outer Galactic disk to the
current position of HE\,0437$-$5439 are depicted in Fig.~4 of E05, who
suggested that the age discrepancy could be overcome if the HVS 
were a blue straggler formed by a merger of less massive stars.
Our revised mass would require a merger of two $\sim$5\,$M_{\sun}$ stars to 
form a blue straggler with the observed parameters for HE\,0437$-$5439.
Their individual life times are about the same as the time of flight,
rendering an origin in the Galaxy very unlikely.

\section{Conclusion}
We have performed a quantitative spectral analysis of the HVS
HE\,0437$-$3954 and reference objects in the LMC and the solar neighbourhood.
Use of state-of-the-art non-LTE modelling techniques allowed precise
elemental abundances to be obtained.

From the results of the spectroscopic analysis we can rule out a GC origin for
HE\,0437$-$3954 because its abundance pattern indicates neither high 
metallicity nor $\alpha$-enrichment. On the other hand a high degree of 
consistency with the LMC pattern is found when accounting for an abundance
scatter within the LMC\footnote{More comprehensive investigations than
presently available, using improved analysis techniques on larger samples of young stars covering
the entire region of the LMC are needed to draw definite
conclusions on the degree of chemical (in)homogeneity in that galaxy.}.
The observed abundance pattern could also be consistent with an origin 
from the outskirts of the Galactic disk, in the case HE\,0437$-$3954 being a
blue straggler. However, a comparison of the time-of-flight 
with the life time of the blue straggler practically rules out this scenario.
Hence, HE\,0437$-$3954 is unlikely of Galactic origin.

\citet{edelmann05} suggested that HE\,0437$-$3954 was ejected from the LMC. Our spectroscopic
analysis lends strong support to this scenario. The abundance pattern is
consistent with that of our LMC reference star. An ejection velocity of about
1000\,{\kms} is required for the star to have reached its position from the LMC.
Accurate proper motion measurements are needed to finally confirm the LMC
origin.

Because no SMBH is known to exist in the LMC, the SMBH slingshot scenario or
other ejection models \citep[e.g.][]{baumgardt06} invoking a SMBH are ruled out.
Hence there must be an additional physical mechanism capable of accelerating
a massive star to a space velocity of 1000\,{\kms}.

Two such alternative scenarios have been proposed.
{\sc i}) A close encounter of a binary with an intermediate-mass black
hole (IMBH) more massive than 10$^3$\,$M_{\sun}$ has been proposed as a viable
ejection mechanism by \citet{GPZ07}, who suggest the populous dense
clusters \object{NGC\,2004} or \object{NGC\,2100}\, in the LMC could harbour
such an~IMBH.
{\sc ii}) Dynamical ejection by interaction of massive binaries in
the cores of dense clusters is plausible \citep{GGPZ08}.
This process is more likely to occur in the LMC than in the Galaxy because the
LMC hosts a significant number of sufficiently massive and dense clusters.
From the list of \citet{mackay03}, eight young clusters can be identified to be 
of the right age to qualify as candidate place of birth of HE\,0437$-$3954. 
Primary candidates are NGC\,2100 and NGC\,2004, followed by \object{NGC\,1850} and
\object{NGC\,1847}. 

In consequence, hyper-velocity stars are not such simple probes
for the shape of the Galactic halo as suggested e.g. by \citet{gnedin05}.
The fundamental assumption of the exclusive origin of HVSs in the GC has to be
dropped, as acceleration may occur in clusters throughout the Galactic disk  
as~well.

\begin{acknowledgements}
We thank the Director General of ESO for making this study possible by granting
director's discretionary time and the staff of the ESO Paranal observatory for their valuable
support. M. F. N. and M. F. gratefully acknowledge financial support by the Deutsche
Forschungsgemeinschaft (grants HE\,1356/44-1 and PR\,685/3-1).
\end{acknowledgements}

\end{document}